\documentstyle[12pt,epsfig]{article}
\input psfig.sty 
\def\reference{\parskip 0pt\par\noindent\hangindent 0.5 truecm}

\begin{document}
%
%

\title{A dynamical model for the CSO-MSO-FRII evolution: hints from hot spot properties}
%

\author{M. Perucho $^{1}$ and J. M.$^{\underline{\mbox{a}}}$ Mart\'{\i} $^{1}$
} 

\date{}
\maketitle

{\center
$^1$ Departamento de Astronom\'{\i}a y Astrof\'{\i}sica, Universidad 
de Valencia.\\ Dr. Moliner 50, 46100 Burjassot (Valencia), Spain\\
e-mail: manuel.perucho@uv.es, jose-maria.marti@uv.es\\[3mm] 
}

%
\begin{abstract}
Compact Symmetric Objects (CSOs) are considered the young counterparts of large 
doubles according to advance speeds measured or inferred from spectral 
ageing. Here we present a simple power law model for the CSO/FRII evolution based on 
the study of sources with well defined hot-spots.
The luminosity of the hot spots is estimated under minimum energy conditions. The advance 
of the source is considered to proceed in ram pressure equilibrium with the 
ambient medium. Finally, we also assume that the jets feeding the hot spots 
are relativistic and have a time dependent power. 
Comparison with observational data allows to interpret the CSO-FRII evolution in terms of 
decreasing jet power with time. 
   
\end{abstract}


\bigskip

%
%

%
%





%
%





\section{Introduction} \label{s:intro}
  
  Compact Symmetric Objects (CSOs) are thought to be early stages of powerful extended radio 
sources as first suggested by Phillips \& Mutel (1982) and later worked out by Readhead et al. (1996).
This view has been underpined by recent measurements 
of hot spot advance speeds (Owsianik \& Conway 1998, Owsianik et al. 1998, Taylor et al. 2000, 
Tschager et al. 2000, Polatidis et al. 2003) and spectral ageing measures (Murgia et al. 2003). 

Current evolutionary models (see Fanti \& Fanti 2002) relate luminosity 
and expansion velocity of a source to jet power and external gas density. The energy 
accumulated in the lobes drives the source expansion. Ram pressure 
equilibrium with the ambient medium is assumed. The volume of the source is usually inferred 
from self-similarity arguments whereas radio power is computed from equipartition 
assumptions. In all the models, jet power was considered as constant. 

  Here we discuss a model, presented in Perucho \& Mart\'{\i} (2002a, Paper II), for 
the long term evolution of powerful radio sources, which is an extension to time dependent
jet power of the model introduced in Perucho \& Mart\'{\i} (2002b, Paper I) for CSOs. 
In order to avoid conjectures about the volume growth of the source (e.g., 
self-similarity), we concentrate in the study of the hot spots for which properties 
like size or luminosity can be derived reliably. In our model we assume that the advance 
work of the hot-spot is directly connected to the jet power. The remaining assumptions of our model are 
standard. The luminosity of the hot spots is estimated under minimum energy 
conditions. The advance of the hot spot proceeds in ram pressure equilibrium with the 
ambient medium. Finally, we also assume that the jets feeding the hot spots are 
relativistic. 

 In section 2, we give the main equations derived from the model. In sectio 3 we present 
the observational data we compared with our model, and section 4 is devoted to 
discussion and conclusions drawn from this work.
  
\section{A dynamical model for the evolution of hot spots in powerful radio sources} 
\label{s:model}

  Our model relies on three basic parameters. The first ($\beta$) is the exponent for 
the growth of linear size of a hot spot with time, $r_{\rm hs} \propto t^{\beta}$ as it
propagates through an external medium with density varying with linear size ($LS$) as 
$\rho_{\rm ext} \propto (LS)^{- \delta}$, where $\delta$ is the second parameter. 

  Using the hot-spot advance speed, we can relate linear size with time and describe the evolution of 
physical parameters in terms of distance to the source of the jets feeding them. Considering that hot-spots
advance with non-relativistic speeds, ram pressure equilibrium leads to,

\begin{equation}
v_{\rm hs} = \sqrt{\frac{F_{\rm j}}{A_{\rm j,hs} \rho_{\rm ext}} }.
\end{equation}

\noindent
where $F_{\rm j}$ is the jet thrust and $A_{\rm j,hs}$, the cross-sectional area 
of the jet at the hot spot, assumed to be $\propto r_{\rm hs}^2$. 
  
  The final step is to consider that for a relativistic jet, jet thrust and
power, $Q_{\rm j}$, are simply related by $F_{\rm j} \approx Q_{\rm j}/c$. If we now allow
for a dependence of the jet power with time like $Q_{\rm j} \propto 
t^{\varepsilon}$ (where $\varepsilon$ is the third parameter), combine all the dependencies and integrate, 
we get

\begin{equation}
v_{\rm hs} \propto (LS)^{\delta/2} t^{\varepsilon/2 - \beta} \rightarrow
t \propto (LS)^{(1 - \delta/2)/(\varepsilon/2 + 1 - \beta)}.
\label{eq:dlsdt}
\end{equation}

\noindent

\noindent
Substituting now in the expressions for the hot spot radius and speed, we obtain

\begin{equation}
r_{\rm hs} \propto (LS)^{\beta(1-\delta/2)/(\varepsilon/2 + 1 - \beta)},
\quad v_{\rm hs} \propto (LS)^{(\delta/2 + \varepsilon/2 - \beta)/
                   (\varepsilon/2 + 1 -\beta)}.
\label{eq:rvhsls}
\end{equation}

  The next equation in our model comes from the source energy balance. We assume 
that the power consumed by the source in the hot spot advance,
$\dot{(PdV)}_{\rm hs,adv}$ adjusts to the evolution of the jet kinetic power, i.e.,

\begin{equation}
\dot{(PdV)}_{\rm hs,adv} (\propto P_{\rm hs} r_{\rm hs}^2 
v_{\rm hs}) \propto Q_{\rm j} \propto t^{\varepsilon}.
\end{equation}

  Finally, under the assumption of minimum energy, the luminosity of 
the hot-spot ($L_{\rm hs}\propto P_{\rm hs}^{7/4} r_{\rm hs}^3 $)
may be expressed in terms of LS 

\begin{equation}
L_{\rm hs}\propto 
(LS)^{\{7/4[-\delta/2(\varepsilon-2\beta) +
\varepsilon/2 -\delta/2 -\beta] +3 \beta(1-\delta/2)\}/(\varepsilon/2-\beta+1)}.
\label{eq:lhsls}
\end{equation}

   Inverting eqs.~(\ref{eq:rvhsls}) and (\ref{eq:lhsls}), we derive expressions
for the evolution parameters of our model in terms of exponents for the evolution of 
observable quantities. 
\begin{equation}
\beta = \frac{s_r}{1-s_v}, \,\,\,\,\, \delta = \frac{12}{7} s_r - \frac{4}{7} s_L + s_v,
\,\,\,\,\, \varepsilon = \frac{2/7 s_r + 4/7 s_L + s_v}{1-s_v},
\label{eq:beta-delta-epsilon}
\end{equation}

\begin{figure}
\begin{center}
\psfig{file=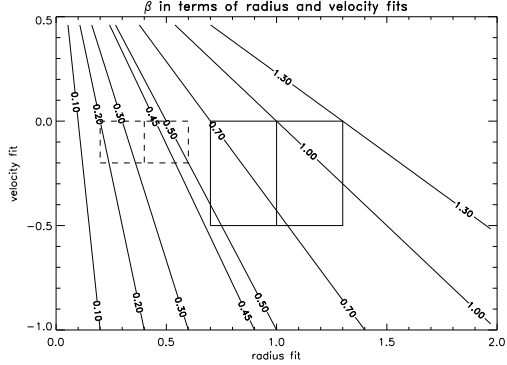,width=7cm,height=5cm} 
\caption{Contours of $\beta$ as function of $s_v$ and $s_r$ for the model discussed in
the text. Boxes bound the expected values of $\beta$ for CSO-MSO evolution (solid lines)
and MSO-FRII evolution (dashed lines).} 
\label{fig:beta}
\end{center}
\end{figure} 

\begin{figure} 
\begin{center}
\psfig{file=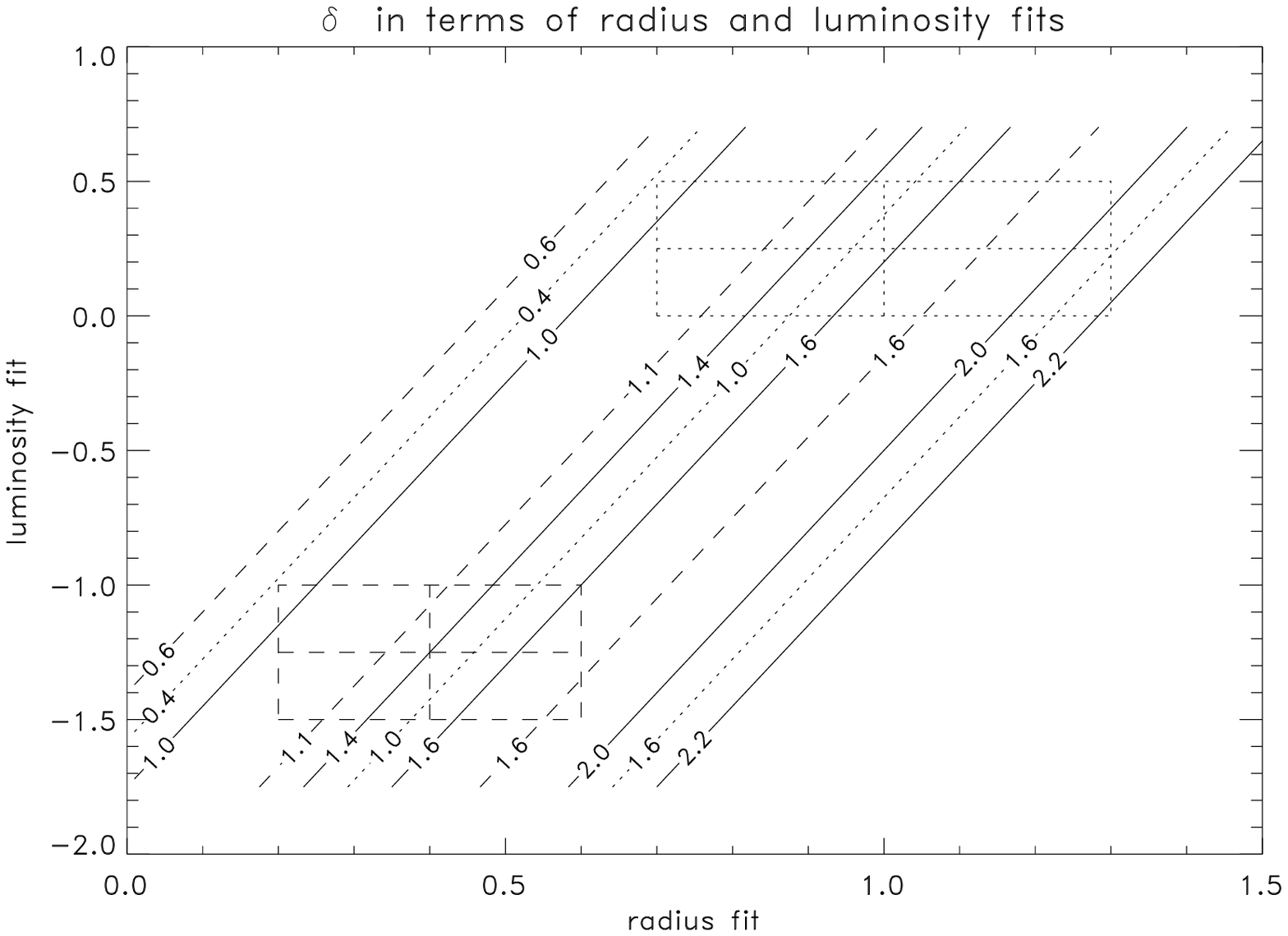,width=7cm} \quad \psfig{file=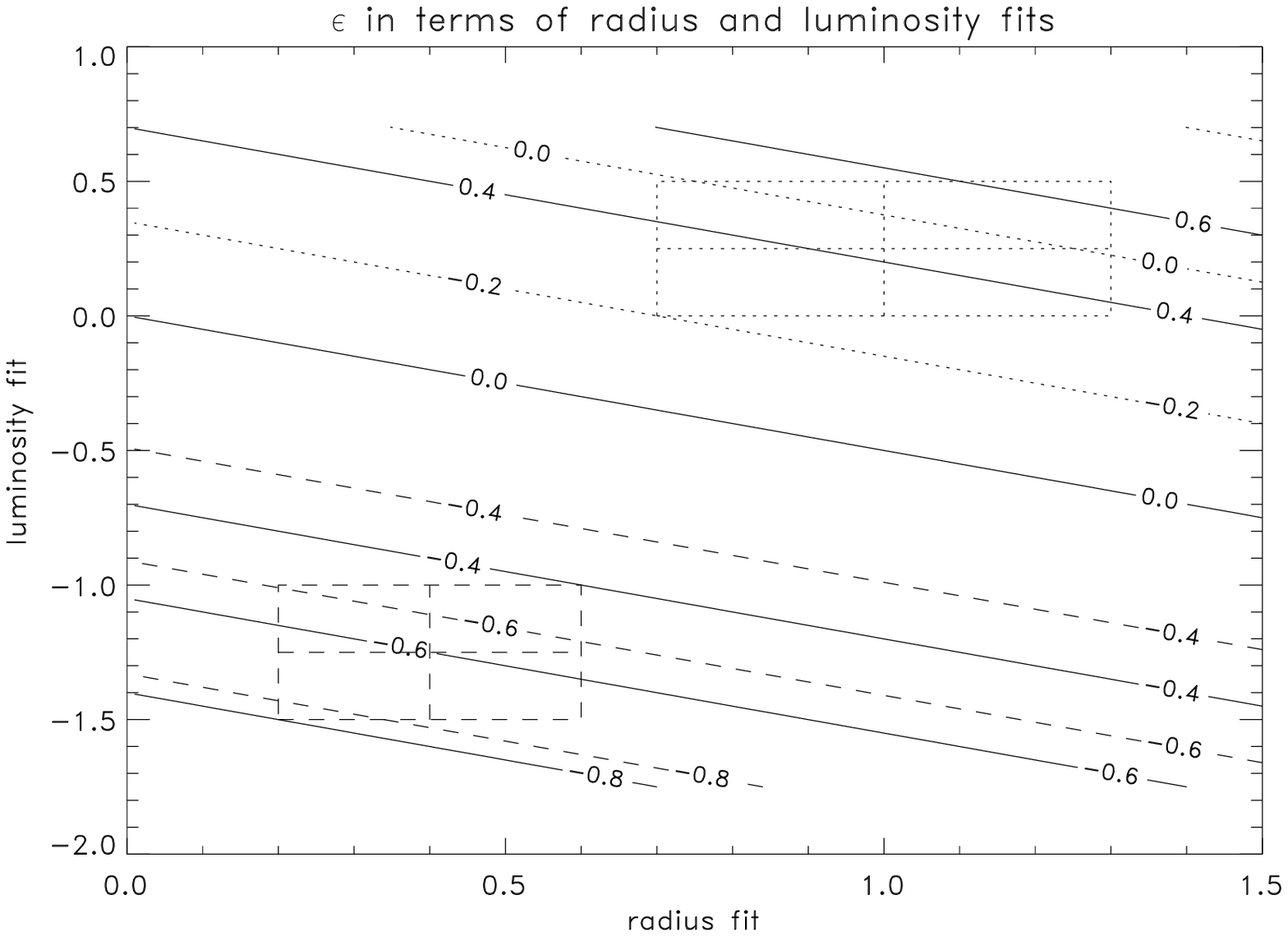,width=7cm}
\caption{Contours of $\delta$ and $\epsilon$ as function of $s_r$ and $s_L$ for the model 
discussed in
the text and different slopes for the hot spot expansion (continuous contours: $s_v = 0$;
dotted contours: $s_v = -0.5$; dashed: $s_v = -0.2$. Boxes bound the expected values of 
$\delta$ and $\epsilon$ for CSO-MSO evolution (dotted lines)
and MSO-FRII evolution (dashed lines).} 
\label{fig:delta-epsilon}
\end{center}
\end{figure}

\noindent
where $s_r$, $s_v$ and $s_L$ stand for the values of the exponents of the observed hot spot radius,
advance velocity and luminosity as functions of $LS$.  
Figure 1 shows the variation of $\beta$ and Figure 2  that of $\delta$ and $\varepsilon$ with the 
different exponents.
Let us note that the previous expressions not only provide a system of
algebraic equations to obtain values for the theoretical parameters in our model.
They also prove that our model is self-consistent since the variations of the 
theoretical parameters induced by those derived from observations agree with physical 
expectations. 
  Therefore, if we are able to obtain values for the exponents in 
eqs.~(\ref{eq:rvhsls}) and (\ref{eq:lhsls}) from observational data
assuming that the corresponding $LS-L_{\rm hs}$ and $LS-r_{\rm hs}$ plots
track the evolution of individual sources, we can derive expected values for $\beta$,
$\delta$ and $\varepsilon$ from the observational fits. 

\section{Observables from hot spots}
\label{s:sample}

  In order to apply our model to the evolution of powerful radio sources from the CSO to
the FRII phases, we have compiled a sample of sources with linear sizes 
between tens of pcs to hundreds of kpcs and well defined hot spots. The sample of CSO is 
the same as the one used in Paper I. Sources were selected from the GPS samples of 
Stanghellini et al. (1997), Snellen et al. (1998, 2000) and Peck \& Taylor (2000). We
have chosen the sources with double morphology already classified in the 
literature as CSOs and also those whose components can be safely interpreted as 
hot spots even though the central core has not been identified yet. The 
criteria we have followed are quite similar to those used by Peck \& Taylor
(2000), although see Paper I for details. Seven {\it Medium size} (1 - 10 kpc) 
symmetric objects (``doubles'') have been taken from the CSS-3CR sample of Fanti et al. 
(1985). Finally, 40 sources from the sample of FRII-3CR radio galaxies of 
Hardcastle et al. (1998) have been considered. See Paper II for further details.

All the subsamples in our combined sample have similar flux density cut-offs: 1 Jy at 5
GHz for the CSOs and 10 Jy at 178 MHz for the MSOs and FRIIs. The differences in redshift
among the three subsamples ($z\leq 1$ for the CSOs; $0.3 < z < 1.6$ for the MSOs; $z<0.3$
for the FRIIs) enhances the luminosity drop between MSOs and FRIIs.

  Figure 3 displays the hot spot radius and 
luminosity with respect to projected linear size on logarithmic scales, 
together with the best linear fit for both CSO-MSO and MSO-FRII subsamples.
Hot-spot linear sizes and luminosities have been calculated as explained in the Appendix of Paper I. 
Table 1 compiles the slopes 
characterizing the power law fits and their errors, as well as the correlation 
coefficients.

\section{Discussion} 
\label{s:disc}

\begin{figure}
\begin{center}
\psfig{file=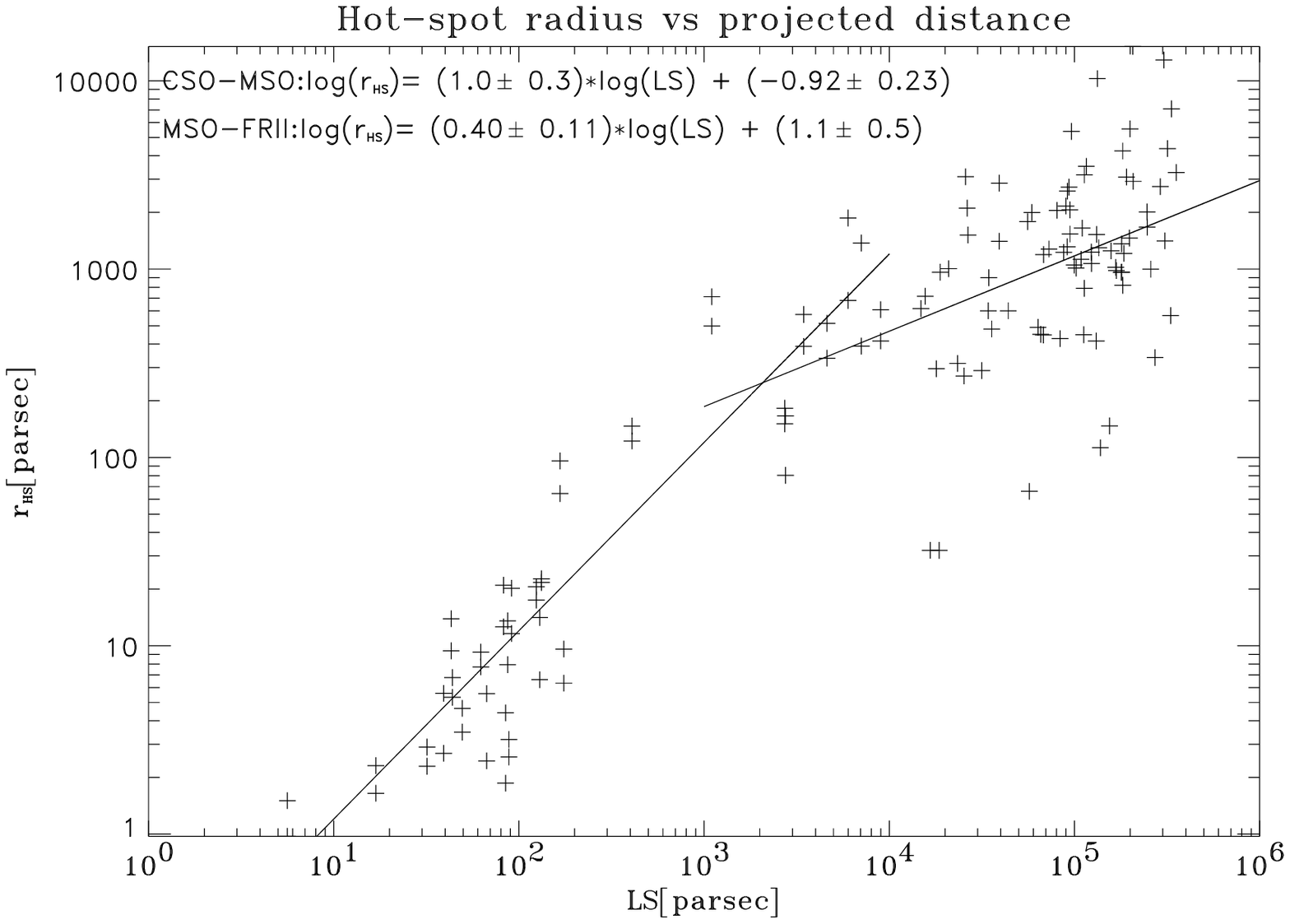,height=5cm} \quad \psfig{file=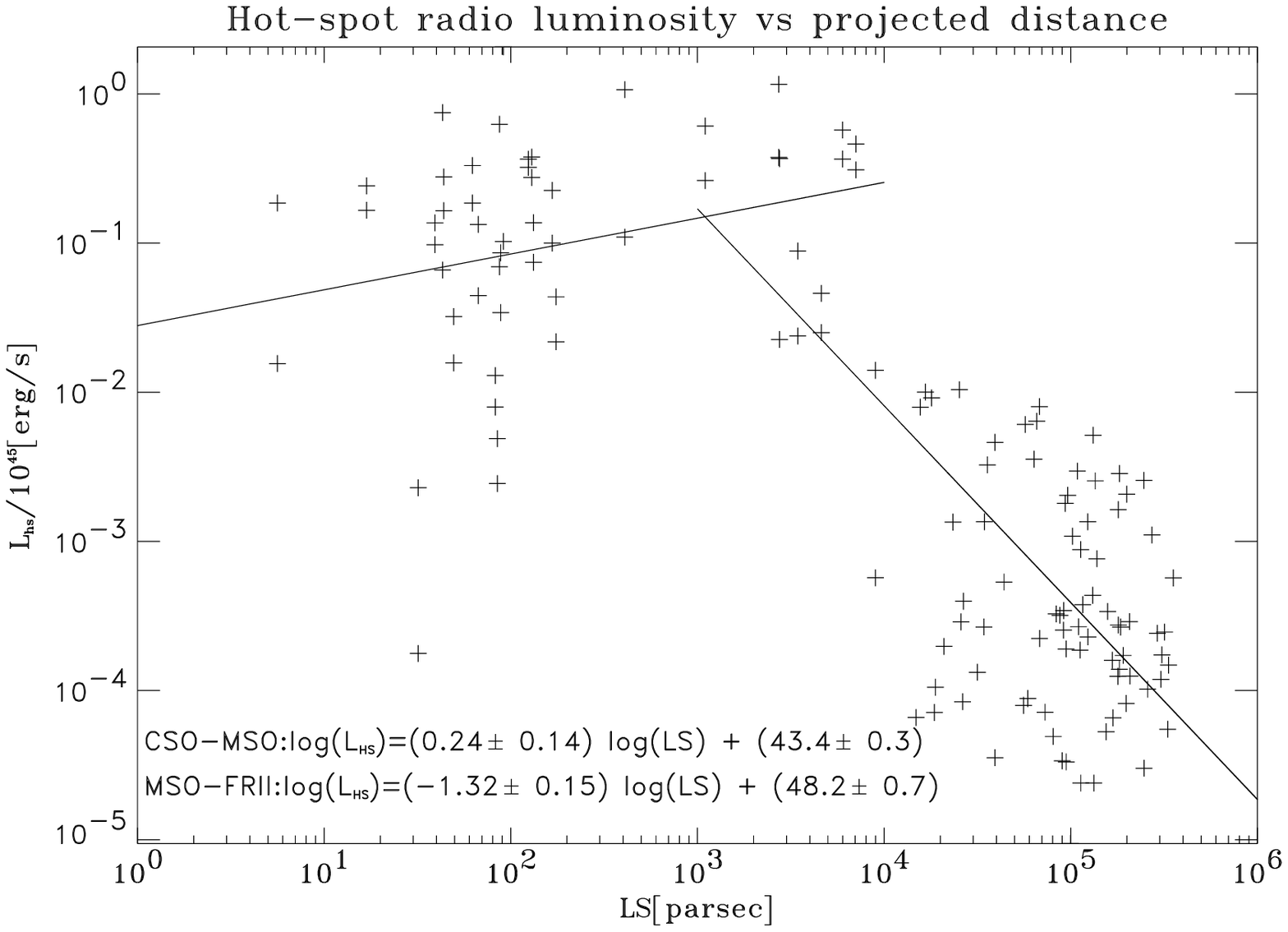,height=5cm}
\caption{Log-log plot for the hot spot radius and luminosity versus projected linear size for the sources in the combined sample.
Continuous lines correspond to the best linear fits for both CSO-MSO and MSO-FRII
subsamples.}
\end{center}
\end{figure} 

  Assuming an evolutionary interpretation of the plots in Fig. 3, the most remarkable 
aspects of those plots are the self-similar growth of hot-spot 
radius with linear size for the first ten kpc of evolution and flattening for large 
sources, in agreement with Jeyakumar \& Saikia (2000), and the change of sign of
the slope for radio-luminosity also at ten kpc. This fact is consistent with the 
transition between the ISM and the IGM, also suggested by the disappearance of IR aligned 
emission after the CSS stage (de Vries et al. 2003), which could imply significant changes
in the evolution.

\begin{table}
\begin{center}
\scriptsize{
\begin{tabular}{c|c c|c c|l|c c c}  \label{t:slopes}
& $r_{hs}$ & & $L_{hs}$ & & Model&  $\beta$ & $\delta$ & $\epsilon$ \\
& $s_r$ & r & $s_L$ & r & & &\\ 
\hline
\hline

&&&&& $s_v = 0$ & $1.0 \pm 0.3$ & $1.6 \pm 0.6$ & $0.4 \pm 0.2$ \\
CSO-MSO&$1.0 \pm 0.3$ & $0.93$ & $0.24 \pm 0.14$& $0.26$ &&&&\\
&&&&& $s_v = -0.5$ & $0.7 \pm 0.2$ & $1.1 \pm 0.7$ & $-0.05 \pm 0.15$ \\
\hline
&&&&& $s_v = 0$ & $0.4 \pm 0.2$ & $1.4 \pm 0.5$ & $-0.6 \pm 0.2$ \\
MSO-FRII&$0.40 \pm 0.11$ & $0.51$ & $-1.32 \pm 0.15$& $-0.69$ &&&&\\
&&&&& $s_v = -0.2$ & $0.3 \pm 0.2$ & $1.2 \pm 0.5$ & $-0.75 \pm 0.15$ \\
\end{tabular}}
\caption{
\footnotesize{Best fits for radius ($s_{r}$) and luminosity ($s_{L}$), along with their 
errors and correlation coefficients ($r$). In the right part of the table we write the 
values of the evolution parameters ($\beta$, $\delta$ and $\epsilon$) which result from the 
calculated best fits and two different possible slopes for advance speed ($s_{v}$), along 
with their errors, which are directly taken from Figs. 1,2 for $\beta$, $\delta$ and $\varepsilon$.  
}}
\end{center}
\end{table}

  Fits can be used to constrain the parameters $\beta$, $\delta$, 
$\varepsilon$ of the model. These values  
appear listed in Table 1. Hot spots undergo a secular deceleration (from 0.2$c$ in the
first kpc to a value $\approx$ ten times smaller in FRIIs) and there are no observational 
indications of any acceleration, hence constant and decreasing hot spots 
velocities have been considered for the CSO-MSO and MSO-FRII fits.

  The break in the fits shown in Fig. 3 at 1-10 kpc produce very
different values for the parameters in the CSO-MSO and MSO-FRII phases. On the contrary, 
there is a complete consistency between the fits for the CSOs alone (see Paper I) and those 
for the CSO-MSO phase. The hot spot expansion rate, $\beta$, decreases from $\approx 1$ in 
the CSO-MSO phase to $\approx 0.4$ in the MSO-FRII phase. Jet power increases with time 
during the first phase ($\varepsilon \approx 0,0.4$ depending on the value chosen for 
$s_v$) and decreases in the long term ($\varepsilon \approx -0.6, -0.7$). It would be 
interesting to relate the time evolution of the jet power with the physical processes 
responsible for the jet production (i.e., accretion rate, black hole spin). The density profile is flat 
($\delta < 2$), consistent for the CSO-MSO stage with that derived by Pihlstroem et al. (2003) 
($\delta \simeq 1.3$) for GPS-CSS sources from HI detections, and the transition between the 
two phases is smooth, although this can be a result of the fitting process that washes out any 
steep gradient between a flat ($\delta \approx 0$), small core and the intergalactic medium. 
We also note that the 
density gradient depends strongly on $s_v$ and that this parameter is poorly known. 
A value of $\delta =2$ in the CSO-MSO phase will produce accelerating hot spots 
($s_v=0.5$) and a large increase of the jet power ($\varepsilon = 2$). Finally, fixing 
$s_L$ and $s_r$ and taking $\delta = 0$, we get $s_v = -1.5$, which is too small to 
be maintained over a long distance: starting with a hot spot speed of 0.2$c$ at 50 pcs,
the hot spot speed at 0.5 
kpc would have decreased up to $6 \, 10^{-3}c$, much smaller than the present accepted 
values for CSO advance speeds (Polatidis et al. 2003, Murgia et al. 2003).  It would be interesting  
to have upper limits of hot spots advance speeds in CSOs in order to constrain the density
profile and the jet power evolution. 

 Regarding the problem of trapped sources, a suitable configuration of external 
medium density and jet power evolution may 
lead to a number of sources which, along with core-jets, may contribute to the excess 
of small sources in count statistics (Marecki et al. 2003, Drake et al. 2003). 

\section*{Acknowledgments}

\footnotesize{
 We thank Prof. R.Fanti for his interest in our work. This research was supported by Spanish 
Ministerio de Ciencia y Tecnolog\'{\i}a (grant AYA2001-3490-C02-01). M.P. thanks the University of 
Valencia for his fellowship within the V SEGLES program and also
thanks LOC and SOC of the 3rd GPS-CSS Workshop for their kindness and hospitality.} 
  
\section*{References}





\reference Drake, C., et al., 2003, PASA, in press 

\reference Fanti, C., Fanti, R., Parma, P., Schilizzi, R.T., \& van Breugel, 
           W.J.M. 1985, A\&A, 143, 292

\reference Fanti, C., \& Fanti, R. 2002, in Issues in Unification of AGNs, eds. Maiolino, R., Marconi, A., Nagar, N. (ASP Conference Series)

\reference Hardcastle, M.J., Alexander, P., Pooley, G.G., \& Riley, J.M. 
           1998, MNRAS, 296, 445

\reference Jeyakumar, S., \& Saikia, D.J. 2000, MNRAS, 311, 397

\reference Marecki, A., et al., 2003, PASA, in press

\reference Murgia, M., et al., 2003, PASA, in press 

\reference Owsianik, I., \& Conway, J.E. 1998, A\&A, 337, 69

\reference Owsianik, I., Conway, J.E., \& Polatidis, A.G. 1998, A\&AL, 336, 37

\reference Peck, A.B., \& Taylor,G.B. 2000, ApJ, 534, 90

\reference Perucho, M., \& Mart\'{\i}, J.M. 2002b, ApJ, 568, 639 (Paper I)

\reference Perucho, M., \& Mart\'{\i}, J.M. 2002a, in preparation (Paper II)

\reference Phillips, R.B., \& Mutel, R.L. 1982, A\&A, 106, 21

\reference Pihlstroem, Y., et al., 2003, PASA, in press

\reference Polatidis, A.G., et al., 2003, PASA, in press

\reference Readhead, A.C.S., Taylor, G.B., Xu, W., Pearson, T.J., Wilkinson, 
           P.N., \& Polatidis, A.G. 1996, ApJ, 460, 612

\reference Snellen, I.A.G., Schilizzi, R.T., de Bruyn, A.G., Miley, G.K., 
           Rengelink, R.B., R\"otgering, H.J.A., Bremer, M.N. 1998, A\&AS, 
           131, 435

\reference Snellen, I.A.G., Schilizzi, R.T., \& van Langevelde, H.J. 2000, 
           MNRAS, 319, 429

\reference Stanghellini, C., O'Dea, C.P., Baum, S.A., Dallacasa, D., Fanti, R., 
           Fanti, C. 1997, A\&A, 325, 943

\reference Taylor, G.B., Marr, J.M., Pearson, T.J., \& Readhead, A.C.S. 2000, 
           ApJ, 541, 112

\reference Tschager, W., Schilizzi, R.T., R\"otgering, H.J.A., Snellen, I.A.G.,
           Miley, G.K. 2000, A\&A, 360, 887

\reference de Vries, W., et al., 2003, PASA, in press

\end{document}